\documentclass[12pt,a4paper]{article}
\usepackage{amssymb,epsf,psfrag}
\begin{document}
\newcommand{\bfcere}{{\bf {\it S. cerevisiae~}}}
\newcommand{\cere}{{\it S. cerevisiae}}
\newcommand{\trans}{{transcription}}
\newcommand{\etal}{{\it et al.}}

\noindent
{\fontfamily{cmss}\fontseries{sbc}\fontsize{22}{1}\selectfont
Identification of essential and functionally moduled genes through 
the microarray assay\\}

\noindent
{\fontfamily{cmss}\fontseries{bx}\fontsize{10}{10}\selectfont
K. Rho$^*$, H. Jeong$^{\dag,\ddag}$ and B. Kahng$^{*,\ddag}$\\}

\noindent
{\sf\fontsize{10}{0}\selectfont  
$^*$School of Physics and Center for Theoretical Physics, \\
Seoul National University, Seoul 151-747, Korea; \\
$^{\dag}$Department of Physics, Korea Advanced Institute of Science and \\
Technology, Daejon 305-701, Korea; and \\
$^{\ddag}$ Department of Physics, University of Notre Dame, \\
Notre Dame, IN 46556, U.S.A.\\}

\vspace{1cm}

\noindent
Identification of essential genes is one of the ultimate 
goals of drug designs. Here we introduce an {\it in silico} 
method to select essential genes through the microarray assay. 
We construct a graph of genes, 
called the gene transcription network, based on the Pearson 
correlation coefficient of the microarray expression level. 
Links are connected between genes following the order of the 
pair-wise correlation coefficients. 
We find that there exist two meaningful fractions of links 
connected, $p_m$ and $p_s$, where the number of clusters becomes 
maximum and the connectivity distribution follows a power law, 
respectively. Interestingly, one of clusters at $p_m$ contains 
a high density of essential genes having almost the same 
functionality. Thus the deletion of all genes belonging to that cluster 
can lead to lethal inviable mutant efficiently. 
Such an essential cluster can be identified in a self-organized way. 
Once we measure the connectivity of each gene at $p_s$. 
Then using the property that the essential genes are likely to 
have more connectivity, we can identify the essential cluster by 
finding the one having the largest mean connectivity per gene 
at $p_m$.
\vfil\eject

\newpage
Thousands of genes and their products in a given living organism is 
believed to function in a concerted way that creates 
the mystery of life~\cite{gavin}. Such cooperative functions 
between genes can be visualized through a graph where nodes denote 
genes and links do activating or repressing effects on transcription
~\cite{marcotte,enright}. 
Traditional methods in molecular biology are very limited to 
analyze such large-scale interactions among thousands of 
genes, so that a global picture of gene functions is hard 
to obtain. 
The recent advent of the microarray assay has enough 
attraction to researcher, allowing them to decipher gene interactions 
in a more efficient way~\cite{kohane}. 
While the data through the microarray assay are not sufficiently 
accumulated yet and they are also susceptible to errors 
in detecting the expression level, the microarray assay is an potential 
candidate for a fundamental approach to understand large-scale gene 
complexes and can be used in many applications such as drug design 
and toxicological research. \\
 
Since the microarray technology is having a significant impact 
on genomics study, many methods for pattern interpretation 
have been developed, including the K-means clustering~\cite{kmeans}, 
the self-organizing map~\cite{som}, the hierarchical 
method~\cite{hier}, the relevance network method~\cite{butte}, etc. 
All such methods, however, contain tunable thresholds, so that  
the results obtained through those methods could be misled 
by the thresholds artificially chosen.  
While those methods are useful for clustering genes, 
they cannot give any information needed to identify essential genes.
Here the essential genes mean target genes for drug designs, 
because the deletion of them leads to inviable mutant to the organism.\\ 

In this paper, we propose a novel {\it in silico} method to identify 
essential genes from microarray database. Our method is inspired by  
the combination of the gene clustering and the close relationship 
between the lethality or essentiality of genes and the connectivity 
in a network. 
Once genes are clustered by using a graph theory and the cluster 
containing a high density of essential genes is identified by using 
the relationship between the lethality and the connectivity of the 
graph~\cite{jeong}. 
The main difference from the previous work~\cite{jeong} 
lies in that while the previous method mainly 
deals with genes individually, our method does rather with clusters 
of genes moduled by their functionality, which turns out to be much 
more efficient in selecting essential genes. 
In our method, we do not use any threshold artificially. 
Thus the essential genes can be identified in a self-organized way. 
Moreover we find that the genes belonging to the same cluster are 
moduled in their functions. Since the essential genes we select 
belong to the same cluster, we can select the essential genes from 
almost the same functional module. 
Finally we propose the functions of unknown genes in the yeast 
protein database classification as the major 
function of the known genes belonging to each cluster.\\
   
\noindent
{\fontfamily{cmss}\fontseries{bx}\fontsize{10}{10}\selectfont
Basic Concepts\\}
The method is inspired from the two concepts: (i) the percolation 
clustering moduled by their functions and (ii) the relationship 
between the essentiality and the inhomogeneous connectivity 
distribution in biochemical networks. 
First the percolation concept~\cite{percolation} is well known 
to physicist and has been applied to many systems including 
the composite system of metals and insulators exhibiting the 
transition as the metal concentration $p$ changes. 
When $p$ is small enough, there are many small size clusters 
of metal and no giant cluster spanning from one end to the other, 
leading the system to be in an insulator phase. 
As $p$ increases, the number of clusters increases. 
There exists a critical value $p_c$ called the percolation threshold 
above which small-size clusters are connected together and a giant 
cluster forms, spanning the entire system. Then the system turns 
into a metal state.\\

Next recently there are many studies for complex 
systems in terms of graph. In graph theory, a graph is 
composed of nodes and links. Degree of a certain node is 
the number of links connected to that node. The emergence 
of a power law in the degree distribution, 
\begin{equation}
P(k)\sim k^{-\gamma},
\label{powerlaw}
\end{equation}
in complex networks has recently attracted many 
attentions~\cite{review1,review2}. 
The network following such a power-law degree distribution is 
called scale-free network.   
The scale-free networks are ubiquitous in nature 
such as social, biological, information systems, etc. 
For example, for the protein interaction network where 
nodes represent proteins and links do their interactions,  
the degree distribution follows a power law~\cite{uetz,ito}. 
Such a behavior implies there exist a few hub proteins 
having a large number of connections compared with other proteins.  
Recently it was shown that such hub proteins are more likely 
to be essential~\cite{jeong}. For the yeast protein interaction 
network, the probability of the proteins with the first 0.7\% ranks is 
as high as 62\%. Thus it was proposed that the selection of 
essential proteins can be made by finding highly connected proteins.\\ 

\noindent
{\fontfamily{cmss}\fontseries{bx}\fontsize{10}{10}\selectfont
Microarray data\\}
We apply those concepts to the microarray data downloaded 
from Ref.\cite{ypd} containing 287 single gene deletion \cere~ mutant 
strains. The deletion data elucidate generic relationships among 
perturbed transcriptomes~\cite{giaever}. 
The data contain two large, internally consistent, global mRNA 
expression subsets of the yeast \cere. 
One of them provides steady state mRNA expression data in wile-type 
\cere~sampled 63 separate times (the `control' set), and the 
other provides individual measurements on the genomic 
expression program of 287 single gene deletion 
mutant \cere~strains grown under identical cell culture 
conditions as wide-type yeast cells (the `perturbation' set). 
Each of the microarray data is the ratio between the expression 
levels of wild-type and perturbed one. Thus the data 
can be written in terms of a $N \times M$ matrix denoted as $\bf C$, 
where $N=6316$ and $M=287$, representing the total number of genes 
and different-deletion experiments, respectively. 
Each element $c_{i,j}$ of the matrix $\bf C$ means the logarithmic 
value (base 10) of the ratio of the expression levels 
for the $i$-th gene under the $j$-th perturbation condition~\cite{hugh}.\\ 

\noindent
{\fontfamily{cmss}\fontseries{bx}\fontsize{10}{10}\selectfont
Percolation clustering \\}
To obtain the correlations among the transcription genes, 
we compare each pair of the expression levels from different genes. 
For each pair, we first select the list 
of genes of which the expression levels are known in both 
transcriptomes. Next the Pearson correlation coefficient 
$\rho_{i,j}$ between $i$ and $j$ genes is calculated, 
defined as 
\begin{equation}
\rho_{i,j}\equiv {{\langle c_{i,k} c_{j,k} \rangle - \langle c_{i,k} 
\rangle \langle c_{j,k} \rangle}\over 
{\sqrt{(\langle c_{i,k}^2\rangle-\langle c_{i,k} \rangle^2)
(\langle c_{j,k}^2 \rangle-\langle c_{j,k} \rangle^2)}}}, 
\label{rhoij}
\end{equation}
where $\langle \cdots \rangle$ means average over $k$, 
different-deletion experiments. 
As shown in Fig.1, the distribution of the correlations 
$\{\rho_{i,j}\}$ is of a bell shape, ranged between -1 and 1. 
Based on the Pearson's coefficients, we generate a network 
by connecting genes whose the Pearson's coefficient is larger than 
a parameter $\rho$. That is, the link between nodes 
$i$ and $j$ is connected if $\rho_{i,j} > \rho$. 
The parameter $\rho$ will be determined later in a self-organized way.
Each link is assumed to have a unit weight. 
Let $p$ mean the fraction of connected links among 
$N(N-1)/2$ possible pairs. Then $p$ depends on $\rho$. 
When $p$ is close to zero ($\rho$ is close to 1), the number 
of links is small, and most nodes remain as isolated nodes 
or form small-size clusters. 
As $p$ increases ($\rho$ decreases), the size of each cluster grows 
or the number of clusters ${\cal N}(p)$ including at least two 
genes increases. At a certain value of $p$, denoted as $p_m$, 
the number of clusters becomes maximum shown in Fig.2, which is  
different from the percolation threshold $p_c$. $p_m$ is estimated 
to be $p_m\approx 0.0002$. Beyond $p_m$, the number of clusters 
decreases, however, the mean size of each cluster increases.\\ 

\noindent
{\fontfamily{cmss}\fontseries{bx}\fontsize{10}{10}\selectfont
Scale-free network \\}
As $p$ increases, the mean size of each cluster increases. 
While the giant cluster forms at $p_c$, 
the degree distribution of the giant cluster does not 
follow a power law. The critical state where the degree  
distribution follows a power law can be reached at a higher fraction 
$p_s\approx 0.0063$ in Fig.3. Note that the degree distribution 
needs an exponential cutoff in the tail part, which is a generic 
behavior due to finite number of genes.       
The degree exponent is estimated to be $\gamma\approx 0.9$, which 
is close to the values obtained by others in different 
systems~\cite{guelzim,provero}, but smaller than typical values 
occurring in many other systems in the range of $2 < \gamma \le 3$. 
For $p > p_s$, the connectivity distribution does not follow 
the power law.\\  
 
To understand the biological implication of the scale-free 
network, we investigate the relationship between the degree of 
a certain gene and its essentiality. 
In Fig.4, we plot the fraction of the essential genes 
with degree larger than $k_{\rm min}$. 
Up to $k_{\rm min}\approx 250$, the genes 
with a larger number of connectivity is more likely to be essential, 
but for $k_{\rm min} > 250$, this tendency does not hold any more. 
Even for the case of $k_{\rm min} < 250$, the fraction of the essentiality 
is not larger than 40\%, less than the rate of 62\% in 
the protein interaction network. Thus as a whole, the way of 
identifying essential genes from the information of the connectivity 
of the gene transcription network alone is not good enough.\\ 

\noindent
{\fontfamily{cmss}\fontseries{bx}\fontsize{10}{10}\selectfont
Method \\}
To improve the success rate of identifying essential genes 
through the microarray assay, here we introduce a new method 
as follows. First links are connected between a pair of genes $\{i,j\}$ 
one by one in descending order of $\rho_{i,j}$. 
Whenever a link is connected, we measure the number of clusters 
${\cal N}(p)$ including at least two genes as a function of $p$. 
Second we identify $p_m$ where the number of clusters becomes 
maximum. Third we find the critical fraction $p_s$ where the 
connectivity distribution follows a power law and measure the 
connectivity of each gene $k_i(p_s)$. 
Finally keeping the information of the degree of each gene 
$k_i$ at $p_s$, we return to the gene transcription network 
at $p_m$. For each cluster $J$, we calculate the average 
connectivity per node, that is,   
\begin{equation}
\langle k^J \rangle=\frac{\sum_{i\in J} k_i^J(p_s)}{N^J(p_m)},
\end{equation}    
where $N^J(p_m)$ is the number of genes belonging to a cluster $J$. 
Based on the fact that genes with a larger number of  
connectivity are more likely to be essential, 
we think that the cluster with the largest value of 
$\langle k^J \rangle$ is the most likely to contain  
essential genes.\\ 

\noindent
{\fontfamily{cmss}\fontseries{bx}\fontsize{10}{10}\selectfont
Essential cluster \\}
To confirm this idea, we directly 
measure the essentiality ${\cal E}^J$, that is 
the fraction of known essential genes among the genes belonging 
to a given cluster $J$. Indeed, as shown in Fig.~5, the 
two quantities, $\langle k^J \rangle$ and ${\cal E}^J$, behave 
in the same manner. Thus we can select the cluster containing 
the largest fraction of essential genes by finding the cluster 
with the largest $\langle k^J \rangle$. 
We find that for the yeast data, 
the third largest cluster with 64 genes turns out to have the 
largest value of $\langle k^J \rangle$, containing 
47 essential genes, 17 nonessential genes, and 
1 unidentified genes (Fig.~6). Thus the certainty of selecting 
essential genes is remarkable improved as high as 73\% 
or even higher when the unidentified gene is excluded, much larger 
than the one obtained only through the connectivity distribution 
in the gene transcription network.\\
  
\noindent
{\fontfamily{cmss}\fontseries{bx}\fontsize{10}{10}\selectfont
Functional clustering \\}
It is known that many biochemical networks are composed of modular 
structure according to their functional role. 
For the yeast, it is known that there are 43 categories by their 
functions~\cite{ypd}. 
We classify genes into 43 categories for each cluster at $p_m$.  
Fig.~7 shows the ratio of genes belonging to each functional 
category for the first five largest clusters. Fig.~8 also shows 
the functional module structure in the gene transcription network. 
From those figures, one can find that there exist major functions 
for each cluster, implying that the genes belonging to the same 
cluster are likely to have the same function. For example, 
the majority of the genes in the largest cluster belong to the 
functional class of amino-acid metabolism. 
Those of the second, third and fourth largest cluster 
are of small molecule transport, RNA processing/modification, 
and protein synthesis, respectively. The reason of such 
functional clustering in the gene transcription network 
lies in that the genes having the same function are likely 
to respond to external perturbation in the same manner, 
making the Pearson correlation coefficients between them large. 
Our result is consistent with the recent discovery of revealing 
modular organization in the yeast transcription network~\cite{ihmels} 
and in the metabolic networks~\cite{rb}. 
Next by using the fact of the gene clustering by their 
functional module, we assign function candidate 
of unknown functional annotation as the major one of the 
genes belonging to the same cluster, which are listed in Table 1.\\ 

\noindent
{\fontfamily{cmss}\fontseries{bx}\fontsize{10}{10}\selectfont
Conclusion and discussion \\}
By using the facts that (i) the genes with the same function are highly 
correlated in the expression level of the microarray and (ii) 
the essential genes are likely to have a larger number of 
connectivity in the large-scale gene transcription network, 
we have proposed an {\it in silico} method to identify a cluster 
containing a high density of essential genes. Since the selected 
genes are from the same cluster, they are likely to be of the same 
function. These essential and functionally moduled genes will be 
useful for drug designs.   
Note that since our method does not include any tuning parameter, it 
has no ambiguity to identify the essential cluster in contrast 
to previous other methods used in gene clustering, where some 
ambiguity is included. Finally our work is similar in idea to a recent 
one that the microarray-driven gene expression can be 
studied much efficiently in parallel to the functional analysis 
of many gene products~\cite{zlauddin}.\\

\noindent
{\fontfamily{cmss}\fontseries{bx}\fontsize{10}{10}\selectfont
Acknowledgements \\}
This work is supported by the ABRL program of the KOSEF. 
The authors would like to thank A.-L. Barab\'asi for helpful 
discussions and hospitality during their visit at the University 
of Notre Dame.\\
\vfil\eject

\vfil\eject

\noindent
Figure Legends \\

\bigskip

\noindent
{\bf Fig.~1: The distribution of the correlation functions}\\

\noindent
The distribution of the Pearson correlation coefficients 
$\rho_{i,j}$ for the yeast \cere.\\ 

\noindent
{\bf Fig.~2: The number of clusters}\\

\noindent
Plot of the number of clusters as a function of the 
fraction $p$ of connected links. \\

\noindent
{\bf Fig.~3: The connectivity distribution of the gene transcription 
network}\\

\noindent
{Plot of the connectivity distribution of the gene transcription 
network at various fractions of link connections, $p=0.0003$ ($\Box$), 
$p=0.0016$ ({\Large$\circ$}), $p=0.0063\approx p_s$ ({\Large$\bullet$}), and 
$p=0.0032$ ({\large$\triangledown$}). At $p_s$, the degree distribution 
follows a power law with an exponential cutoff. Dotted line having 
a slope -0.9 is drawn for the eye.}\\ 

\noindent
{\bf Fig.~4: The fraction of the essentiality}\\

\noindent
{Plot of the fraction of the essentiality of nodes 
having degree larger than $k_{\rm min}$ as a function of 
$k_{\rm min}$}\\

\noindent
{\bf Fig.~5: The identification of the essential cluster}\\

\noindent
{The comparison between $\langle k^J \rangle$ ({\Large$\bullet$}) 
and ${\cal E}^J$ ($\Box$) for each cluster indexed by 
cluster size at $p_m$.}\\ 

\noindent
{\bf Fig.~6: The gene transcription network colored by their 
essentiality}\\

\noindent
{The gene transcription network at $p_m$ of the 
yeast \cere. The green, white and yellow nodes represent essential, 
nonessential, and unidentified genes, respectively.}\\

\noindent
{\bf Fig.~7: The functional genes ratio for each cluster}\\

\noindent
{The genes ratio belonging to each functional category for 
the first five largest clusters.}\\ 

\noindent
{\bf Fig.~8: The gene transcription network colored by their 
functions.}\\

\noindent
{The gene transcription network at $p_m$ of the 
yeast \cere. The genes with the functions, amino-acid metabolism, 
small molecule transport, RNA processing/modification, protein 
synthesis are distinguished by the different colors, 
red, blue, green, and brown, respectively. The white and 
the yellow represent other functions and unknown function, 
respectively.}\\ 

\vfil\eject

\noindent
Table Legend \\

\noindent
{Table 1: Function candidates for unknown genes} \\ 

\noindent 
{Assigned functions for unknown genes by following the 
major function of the genes of each cluster at $p_m$.}\\

\vfil\eject

\begin{figure}
\psfrag{rho}{\large {$\rho$}}
\psfrag{P(rho)}{\large {$P$($\rho$)}}
\centerline{\epsfxsize=12cm \epsfbox{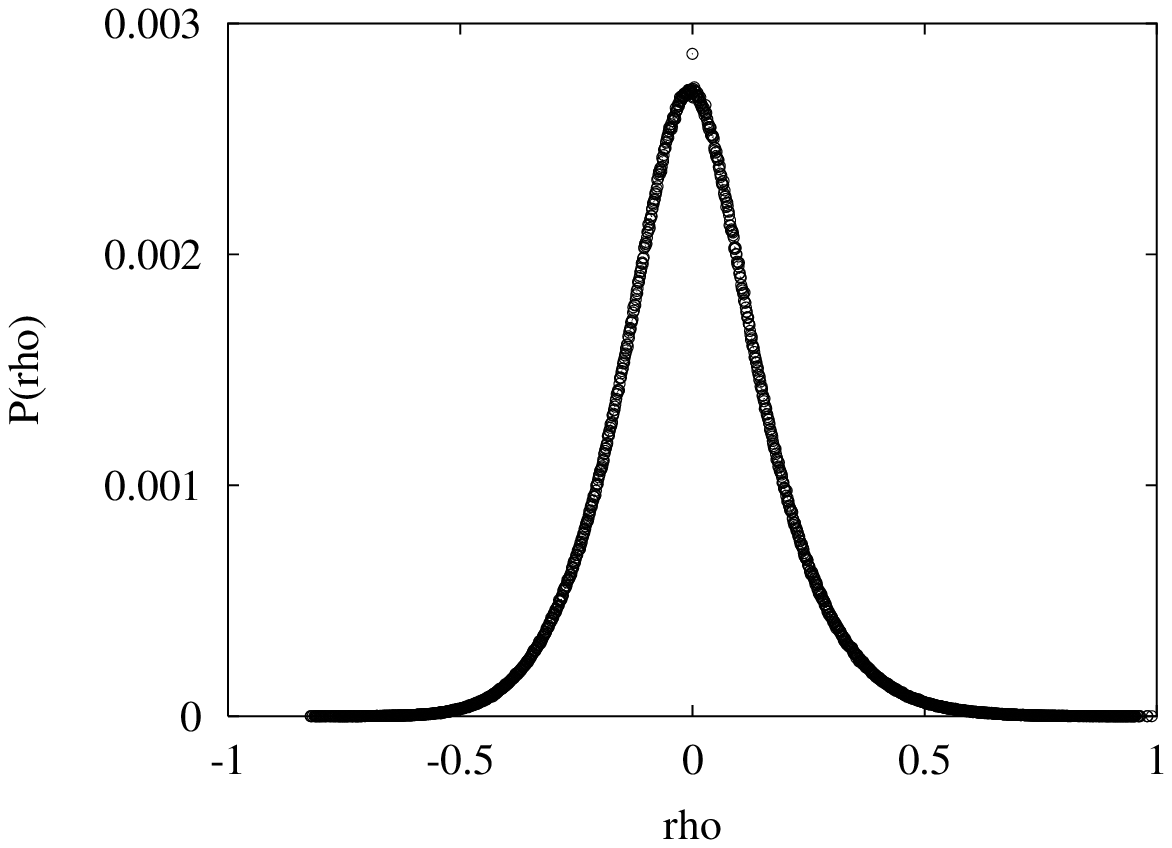}}
\caption{Rho \etal}
\label{dist} 
\end{figure}

{\hbox{}}
\vfil
\eject

\begin{figure}
\psfrag{N(p)}{\large {${\cal N}(p)$}}
\centerline{\epsfxsize=12cm \epsfbox{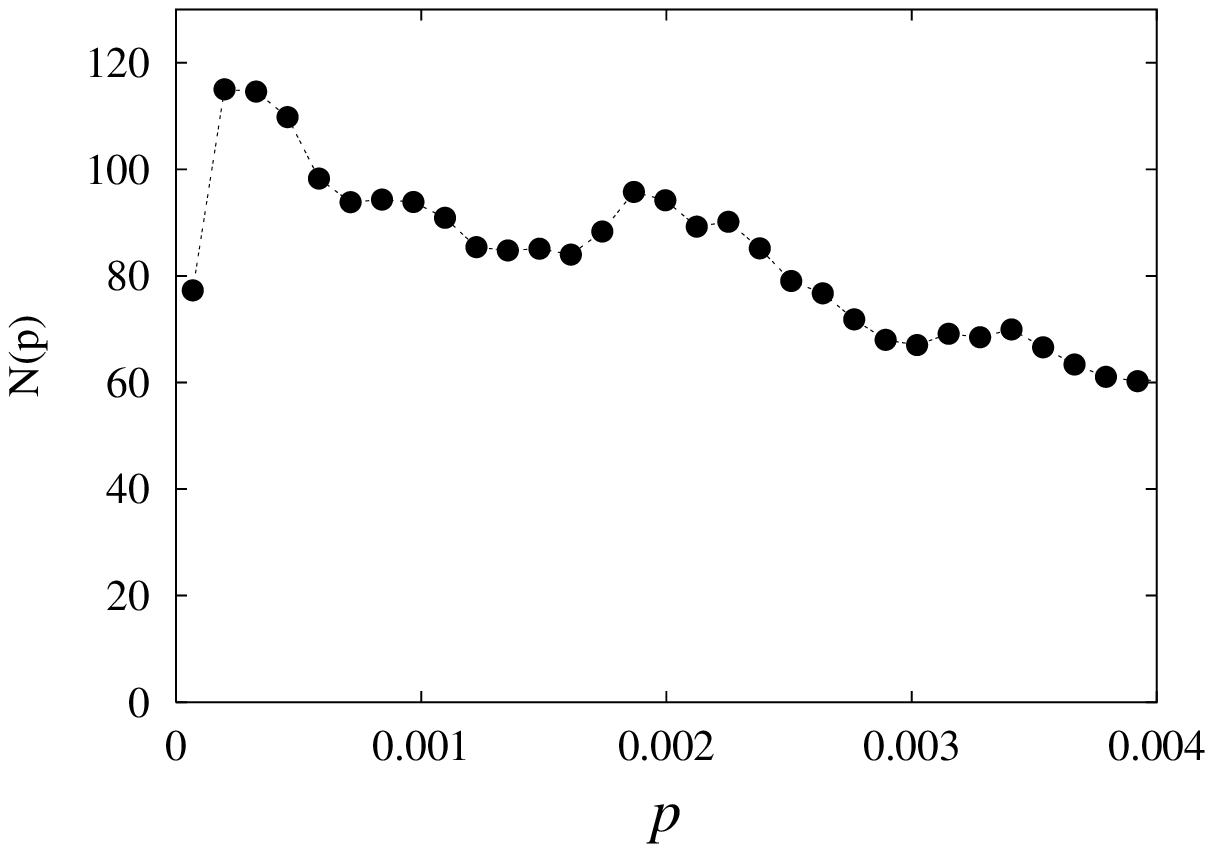}}
\caption{Rho \etal}
\label{n_cluster} 
\end{figure}
{\hbox{}}
\vfil\eject

\begin{figure}
\centerline{\epsfxsize=12cm \epsfbox{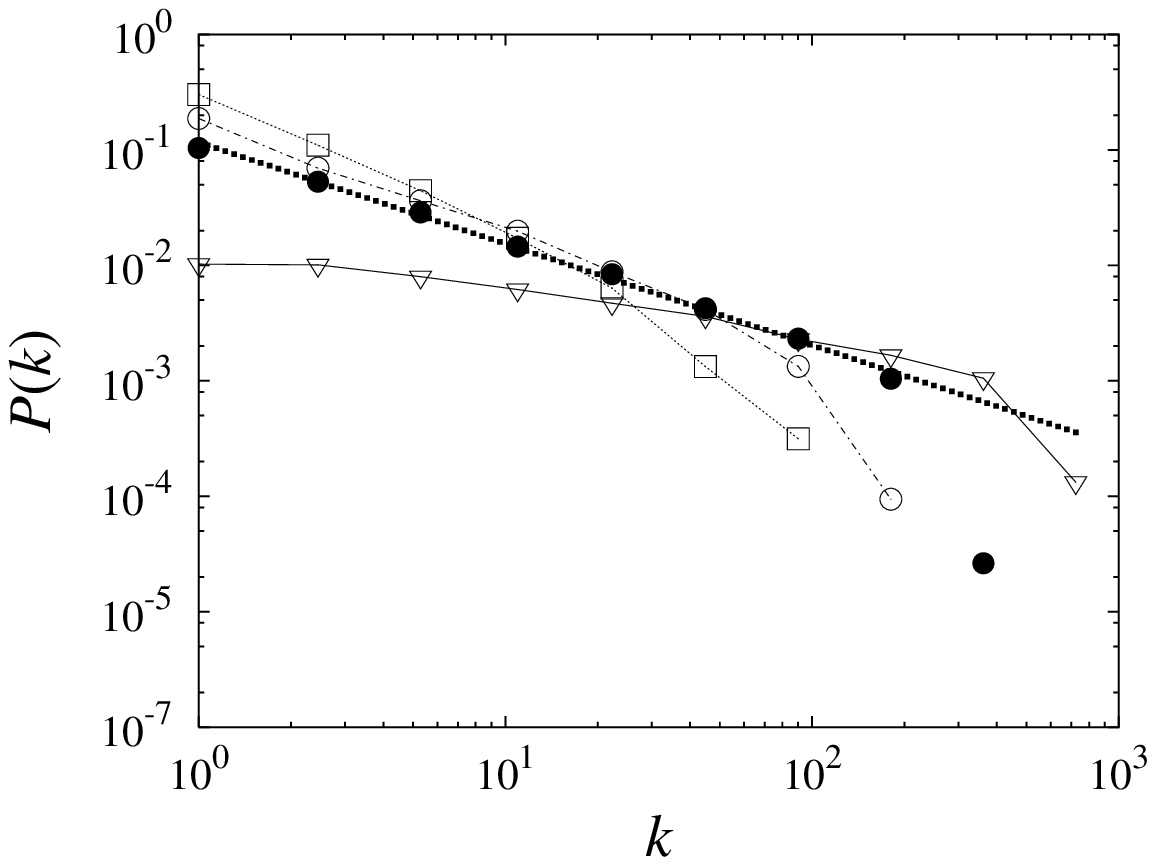}}
\caption{Rho \etal}
\label{sf} 
\end{figure}
{\hbox{}}
\vfil\eject

\begin{figure}
\centerline{\epsfxsize=12cm \epsfbox{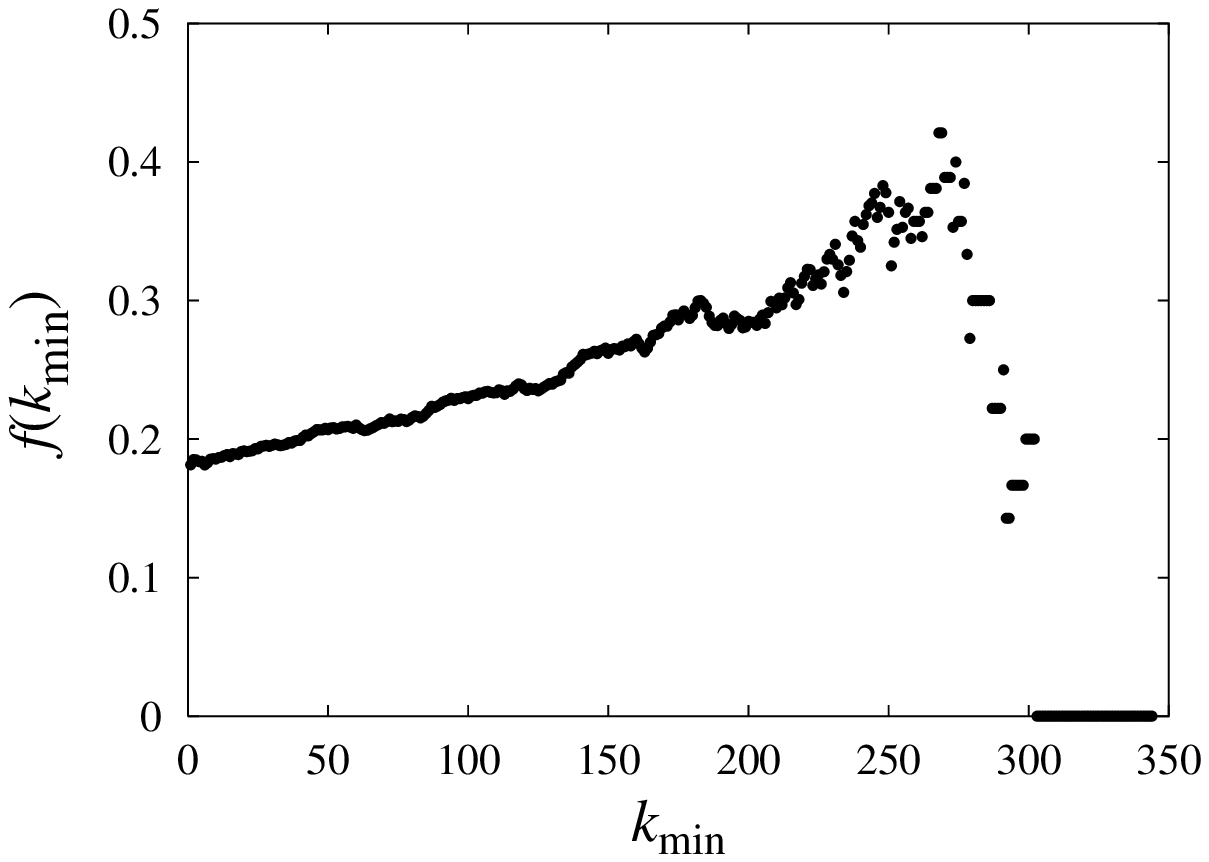}}
\caption{Rho \etal}
\label{essential} 
\end{figure}
{\hbox{}}
\vfil\eject

\begin{figure}
\psfrag{EcaretJ}{\large {${\cal E}^J$}}
\centerline{\epsfxsize=12cm \epsfbox{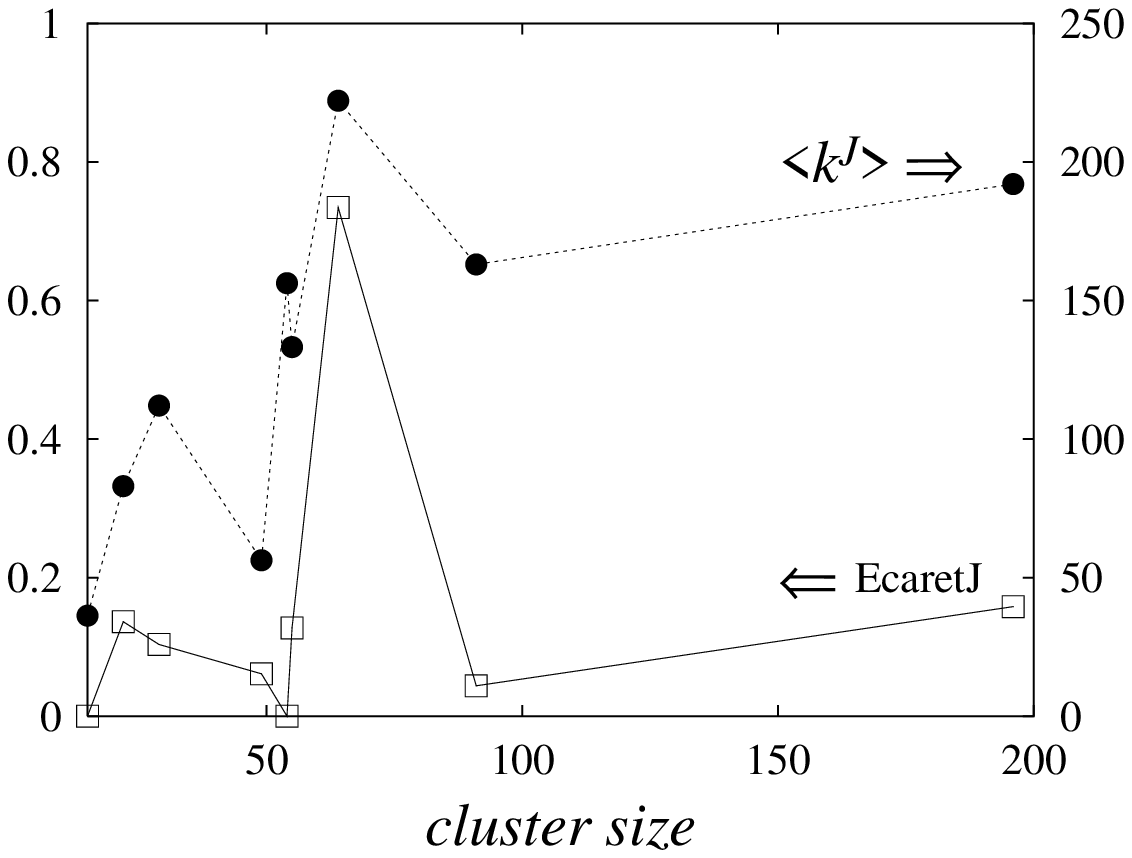}}
\caption{Rho \etal}
\label{comparison} 
\end{figure}
\vfil\eject

\begin{figure}
\centerline{\epsfxsize=12cm \epsfbox{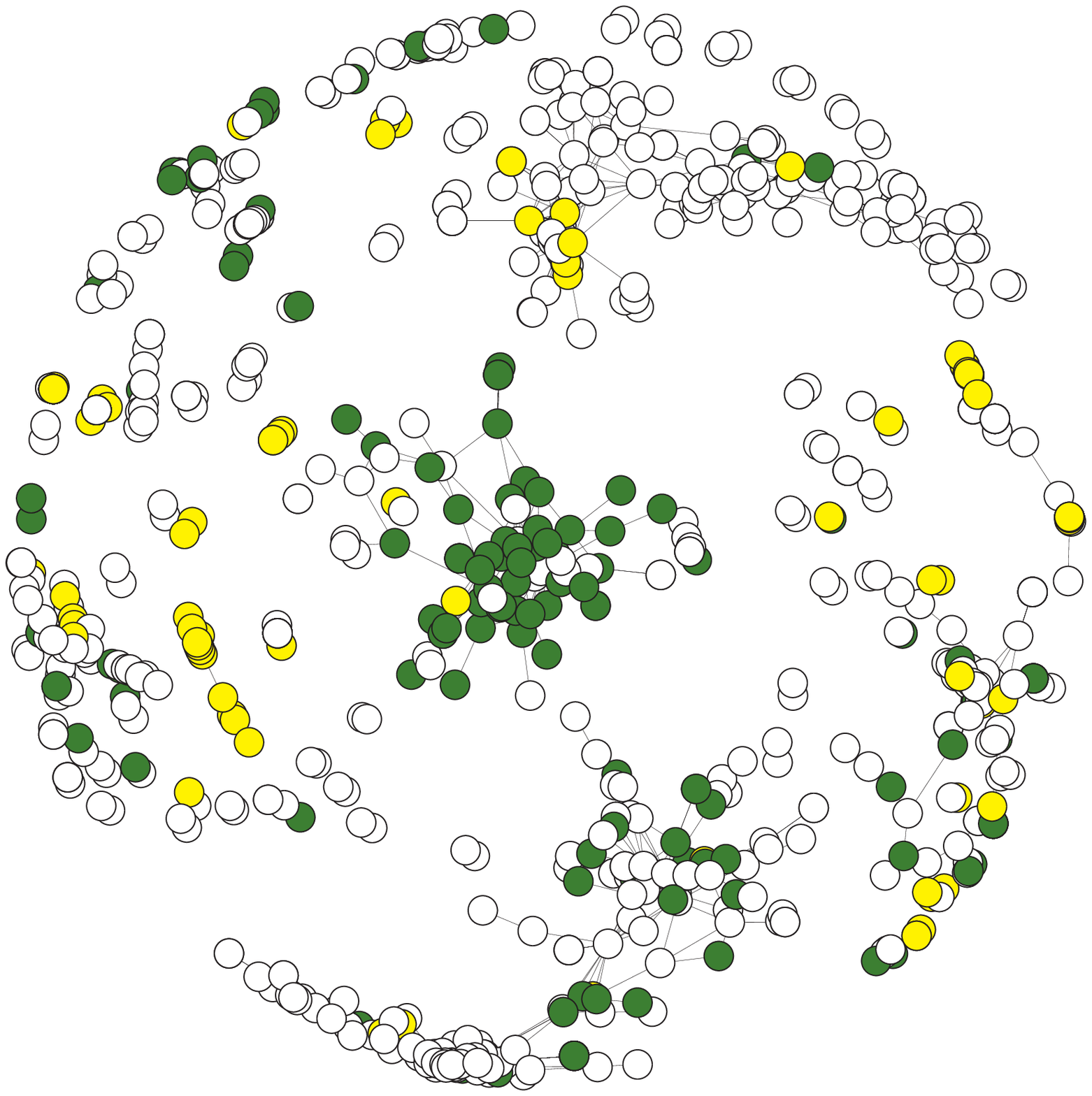}}
\caption{Rho \etal}
\label{0.5essential}
\end{figure}
\vfil\eject

\begin{figure}
\centerline{\epsfxsize=12cm \epsfbox{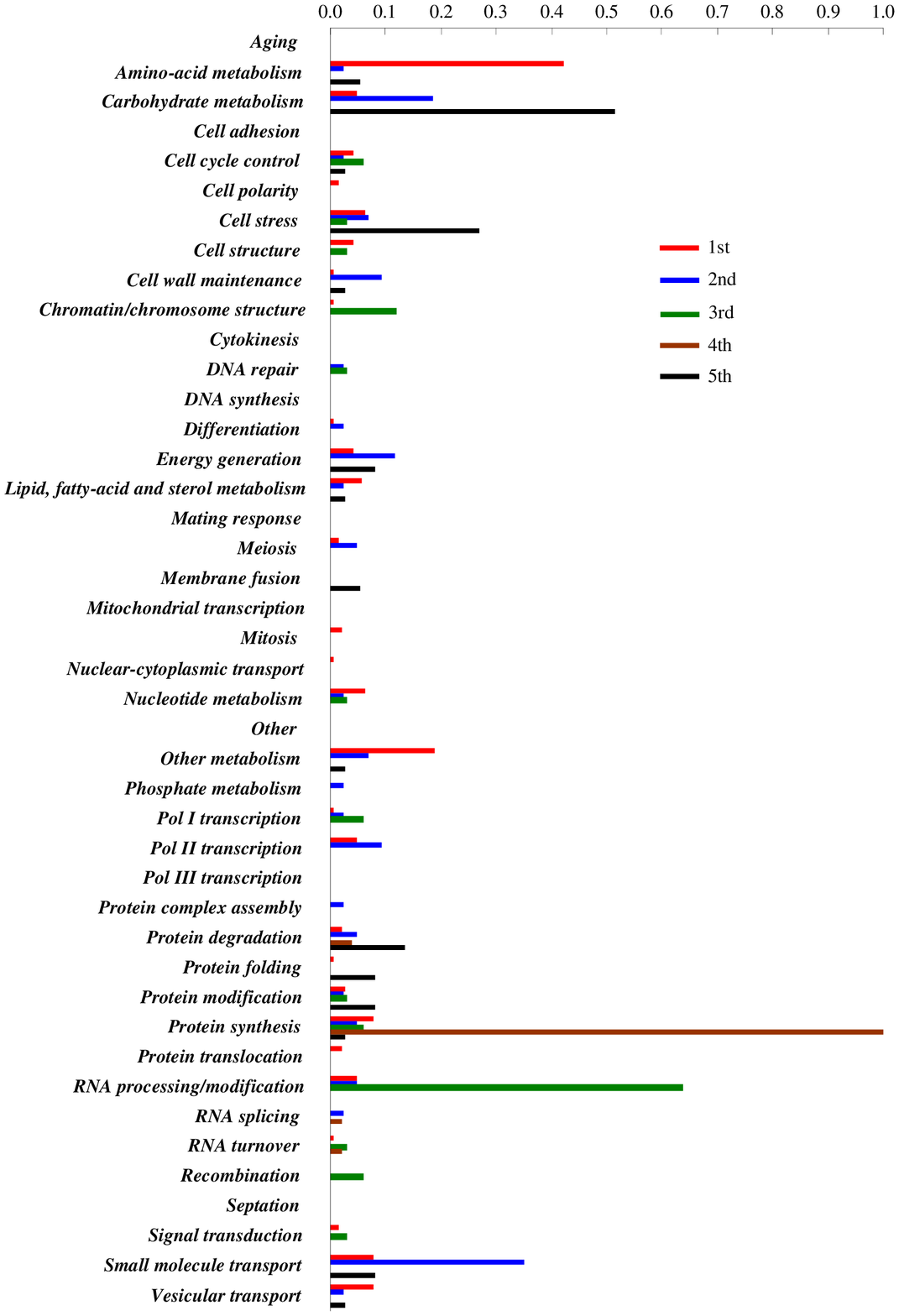}}
\caption{Rho \etal}
\label{histogram}
\end{figure}
\vfil\eject

\begin{figure}
\centerline{\epsfxsize=12cm \epsfbox{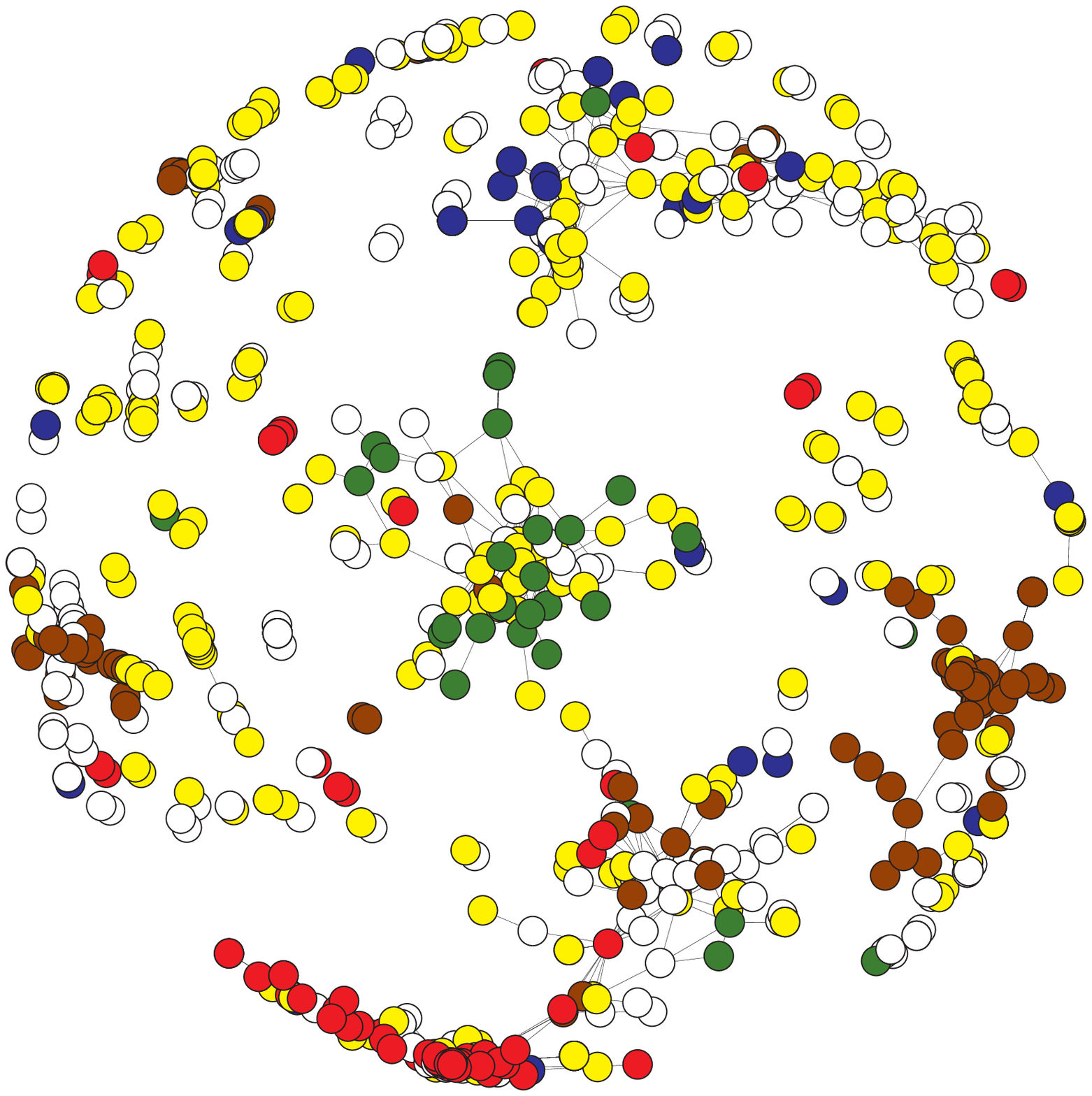}}
\caption{Rho \etal}
\label{0.5modularity}
\end{figure}
\vfil\eject

\begin{table}
\caption{Rho \etal}
{\footnotesize
\begin{tabular}{|c|c|c|c|c|c|c|} \hline
\multicolumn{7}{|l|}
{Amino-acid metabolism or Other metabolism - 52 genes}\\
 \hline \hline
YAL014C&  YBR046C&  YBR047W&  YBR147W&  YBR261C&  YCL028W&  YCL044C\\ \hline
YCR051W&  YDL054C&  YDR425W&  YDR426C&  YER152C&  YER175C&  YFL010C\\ \hline
YFL028C&  YGL117W&  YGL224C&  YHR029C&  YHR122W&  YHR162W&  YIL041W\\ \hline
YIL056W&  YIL164C&  YIL165C&  YJL072C&  YJL200C&  YJL213W&  YJR111C\\ \hline
YJR130C&  YJR154W&  YLR152C&  YLR193C&  YLR267W&  YLR290C&  YLR339C\\ \hline
YML113W&  YMR097C&  YMR321C&  YNL129W&  YNL276C&  YNL311C&  YNR069C\\ \hline
YOL118C&  YOR042W&  YOR044W&  YOR203W&  YPL135W&  YPL251W&  YPL264C\\ \hline
YPR059C&  YPR114W&  YKL033W-A&&&&\\ \hline

\multicolumn{7}{l}{}\\ \hline

\multicolumn{7}{|l|}
{Small molecule transport - 48 genes}\\
 \hline \hline
YAL065C&  YBR301W&  YCR104W&  YDR542W&  YEL049W&  YER188W&  YFL020C\\ \hline
YGL261C&  YGR150C&  YGR169C&  YGR294W&  YHL046C&  YHR049W&  YIL176C\\ \hline
YIR041W&  YJL218W&  YJL223C&  YKL005C&  YKL224C&  YLL025W&  YLL056C\\ \hline
YLL064C&  YLR037C&  YLR091W&  YLR269C&  YLR461W&  YMR020W&  YMR107W\\ \hline
YMR252C&  YMR253C&  YNL285W&  YNL310C&  YNR014W&  YNR076W&  YOL161C\\ \hline
YOR134W&  YOR205C&  YOR286W&  YOR389W&  YOR394W&  YPL107W&  YPL277C\\ \hline
YPL282C&  YPR053C&  YAL068C&  YHR049C-A&  YMR316C-B&  YMR325W&\\ \hline

\multicolumn{7}{l}{}\\ \hline

\multicolumn{7}{|l|}
{RNA Processing/modification - 31 genes}\\
 \hline \hline
YBL028C&  YCL059C&  YDL063C&  YDL148C&  YDR101C&  YDR152W&  YDR165W\\ \hline
YDR324C&  YDR361C&  YDR496C&  YER126C&  YGR128C&  YGR145W&  YHR052W\\ \hline
YHR085W&  YHR196W&  YHR197W&  YKR060W&  YKR081C&  YLR068W&  YLR129W\\ \hline
YML093W&  YNL002C&  YNL182C&  YNL207W&  YNR053C&  YOR004W&  YOL077C\\ \hline
YOR145C&  YPL012W&  YPL146C&&&&\\ \hline

\multicolumn{7}{l}{}\\ \hline

\multicolumn{7}{|l|}
{Protein Synthesis - 5 genes}\\
 \hline \hline
YGL102C&  YJL188C&  YLR062C&  YPL142C&  YPR044C&&\\ \hline

\multicolumn{7}{l}{}\\ \hline

\multicolumn{7}{|l|}
{Carbohydrate metabolism or Cell stress - 17 genes}\\
 \hline \hline
YBR053C&  YDL110C&  YDL204W&  YDR032C&  YER067W&  YIL136W&  YJL070C\\ \hline
YJL161W&  YLR149C&  YLR270W&  YML128C&  YMR110C&  YNL115C&  YNL200C\\ \hline
YOL082W&  YPL123C&  YMR169C&&&&\\ \hline

\multicolumn{7}{l}{}\\ \hline

\multicolumn{7}{|l|}
{Energy generation - 7 genes}\\
 \hline \hline
YGL069C&  YKL169C&  YKL195W&  YMR158W&  YPR099C&  YPR100W&  YKL053C-A\\ \hline

\multicolumn{7}{l}{}\\ \hline

\multicolumn{7}{|l|}
{Chromatin/chromosome structure - 3 genes}\\
 \hline \hline
YBL113C&  YFL068W&  YML133C&&&&\\ \hline

	\end{tabular}
}
\label{candidate}
\end{table}

\end{document}